\documentstyle[12pt]{article}
\begin{document}
\textwidth 160mm
\textheight 240mm
\topmargin -20mm
\oddsidemargin 0pt
\evensidemargin 0pt
\newcommand{\beq}{\begin{equation}}
\newcommand{\eeq}{\end{equation}}
\begin{titlepage}
\begin{center}

{\bf Threshold amplitudes in  field theories  and integrable systems}

\vspace{1.5cm}

{\bf  A.Gorsky,K.Selivanov }

\vspace{1.0cm}

{ITEP,B.Cheryomushkinskaya 25,Moscow,Russia}

\vspace{1.9cm}

{ITEP-61/95}

\vspace{1.0cm}

\end{center}

%\begin{document}
%\maketitle
%vspace{3.5 cm}

\begin{abstract}
We discuss the threshold tree amplitudes in diverse nonintegrable
quantum field theories in the framework of integrability. The
amplitudes are related to some Baker functions defined on the auxiliary
spectral curves and the
nullification phenomena are shown to allow a topological interpretation.
\end{abstract}
%\vfill
\end{titlepage}

%\begin{document}
\newpage
\setcounter{equation}{0}
1.Threshold amplitudes and formfactors have been intensively discussed
recently. The main task is to understand
their dependence on the number of final particles n  when n is large. As a
byproduct, an intriguing nullification phenomenon has been observed
\cite{voloshin2,argyres2,smith}, namely, in a number of theories the threshold
amplitudes in the tree approximation with two initial and n final particles
appeared to be zero for n bigger then some model-dependent number - 2,3,4...
-  implying thus
a sort of hidden integrability in the theory.

This  phenomena of hidden integrability  in  some  sector  of
nonintegrable field theory is the subject of this letter.The first
step in this direction was done in  \cite{rubakov}  where in some
particular theory the nullification was related to some additional
symmetry of the model which in turn existed due to the fact that the model on
the threshold is an integrable one (so called Garnier model).
On the other hand,  in  the  theory  of  integrable
systems a classical solution  (to be precise, classical  $\tau$
function) of one integrable system often happens  to  be  a  generating
function for the quantum amplitudes in the other one. As examples one
can mention relation  between  $\tau$-function of Sin-Gordon model
and Ising correlators \cite{korepin} or the relation between  Toda
lattice Baker function and tachionic scattering amplitudes  in the
string theory  \cite{moore}.

Here  we  show  how the threshold
amplitudes  in  nonintegrable  4D  theories  fall  into  the  same
scheme. The Baker function of the particular  finite-gap  solution of
the relevant integrable field theory is  the  generating  function
of the threshold amplitudes in a sence that the latter are residues of
the former at its poles on the auxiliary spectral curve. The nullification
appeares to be a consequence of  the  Riemann-Roch  theorem  which
states that
the number of poles of the Baker function (zeros of the $\tau$-function)
equals to the genus
of the spectral curve.

2.According to the general principles of the quantum field theory,
(see also \cite{voloshin},\cite{brown})
the generating function for the connected parts of the tree amplitudes
is given by a solution of the
classical equation of motion ${\phi}(t,x;a,{a}^{*})$ :

\beq
\label{classical}
\sum<1_{k};a_{p_{1}}...|a^{*}_{q_{1}}...>\frac{a^{n}a^{*m}}{n!m!}=\\
\int e^{-ikx}(\partial^{2} +m^{2})\phi(x,t)
\eeq
and the solution $\phi$ is fixed by the Feynman-type condition
\beq
\phi=\sum_{q}(a_{q}e^{-iqx}+a_{q}^{*}e^{iqx}) +O(\lambda)
\eeq
where $\lambda$ is coupling constant of the theory. In general the
only way to find this solution is to perform
the perturbation expansion of the
equation what is more or less parallel to summing the tree Feynman
graphs. To proceed in a different way one has to reduce to  the
threshold kinematics.
The equations of motion then become ordinary differential equations and have
more chances to be solvable. In some cases -  in  fact,  not  very
exceptional
ones - the reduced equations happen to coincide with equations of motion
of some integrable model. Those are basically the cases when the
nullifications take place for all particles on the threshold.

Note that in the formula (\ref{classical}) the particle with momentum $k$ does
not enter the asymptotic condition and thus can be left off the threshold.
To have another particle off the threshold one takes a variation of
formula \ref{classical} with respect to, say, $a_{p}$ then puts $a_{p}$
equal to zero,
and finally  reduces to the  threshold.  Therefore   the  formula  for
the amplitude takes the following form
\beq
\sum <1_{\omega,\vec k}1_{\omega,\vec -k}...|...>\frac{a^{n}a^{*m}}
{n!m!}=\int e^{-i\omega_{k} t}(\partial_{t}^{2}+k^{2}+m^{2})\Psi_{k}
\eeq
where
\beq
\Psi_{k}(t;a,a^{*})e^{-i{\vec k}{\vec x}}
=\frac{{\delta}{\phi(t,x;a,a^{*})}}{{\delta}{a_{k}}}.
\eeq
It   is    easily    seen    that $\Psi(t;a,{a}^{*})$
obeys the equation
\beq
L\Psi=0
\eeq
with some linear operator L and the asymptotic condition
\beq
\Psi=e^{-i\omega t}+O(\lambda)
\eeq
thus being a kind of the Baker function. In this situation the nullification
takes place when the operator L happens to be a finite gap (or degenerated
finite gap) Lax operator for some integrable system. The energy of the
in-particle $\omega_{k}$ plays the role of the spectral parameter and
the amplitudes are nothing but the residues of the
Baker function  $\Psi(t;a,{a}^{*})$ on the spectral curve.

Further we consider some examples of this general picture.

3.Let us start with the simplest kinematical region when both initial
and final particles are at rest. We restrict  ourselves  by  two
examples. At first consider the massive O(3) $\sigma$ model
with the Lagrangian
\beq
L={(\partial{\phi})}^2+\sum m_{i}^{2}{\phi_{i}}^{2}+
{\lambda}({\phi}^{2}-{\eta}^{2})
\eeq
where $\lambda$ is Lagrangian multiplier.This Lagrangian reduces to
the Neumann system describing the motion of particle on the  sphere
under the action of the linear force.This system  belongs  to  the
class of the ones whose flows are linearized on the Jacobian  of  the
spectral curve defined in terms of the Lax operator
\beq
det(L(z)-\rho)=0
\eeq

The solution of the equation of motion is as follows \cite{dubrovin}
\beq
x_{1}=\beta_{1}\frac{\theta_{0,\frac{1}{2},0,0}(t\vec U+z_{0})
\theta(z_{0})}{\theta_{0,\frac{1}{2},0,0}(z_{0})\theta(t\vec U+z_{0})}
\eeq
where $\beta_{1}^{-2}=\prod_{i\neq 1}(m_{1}-m_{i})$, $z_{0}$ is the
point of the Jacobian, and $\vec U$ is the standard vector whose
components are the integrals of the normalized holomorphic differentials along
A-cycles. The expressions for another $x_{i}$ are analogous and
then to get the appropriate solution one needs
to fix  time boundary conditions.Note that according to  the
standard arguments integrals of motion are defined as
\beq
H_{k}=Tr{L}^{k}
\eeq

To  include  Baker  function  into  the   game remind the relation
between the Neumann dynamics and the flows in the space of
two-gap KdV  potentials  \cite{moser}.It  can  be  formulated  as
follows: if we introduce hyperelliptic spectral curve
\beq
y^{2}=\prod_{i=1}^{5}(x-x_{i})
\eeq
with $x_{2i+1}=-m_{i}^{2},i=0,1,2$ and consider the Baker  function
$\Psi$  for  KdV  solution,  the
following relation holds true
\beq
\beta_{i}x_{i}(t)=\Psi(z,m_{i}^{2})
\eeq
Therefore the investigation of singularities of  $x_{i}(t)$  as  a
function  of  $m_{i}$  is  equivalent  to  the  search   for   the
singularities  of  the  Baker  function  when   branching   points
coincide.Residues  of  the  Baker  function  at  the  points   of
singularities give nonzero amplitudes in the bare theory.

Let us turn to another simple example of  the  same  nature,namely
O(N) ${\phi}^{4}$ model with the action

\beq
L=\sum_{i}(\partial{\phi}_{i})^{2}+\sum_{i}{m_{i}^{2}\phi_{i}}^{2}+
{\lambda}(\sum_{i}{\phi_{i}}^{2})^{2}
\eeq

Zero  mode  model  is  now  well known  Garnier  system  describing
geodesic motion on the symmetric  spaces which  again  is  related
to N-gap KdV solution \cite{choodnovsky} as follows
\beq
u(x)=\sum_{i=1}^{N}x_{i}^{2}
\eeq

Then it is possible to connect  Garnier  particles
with the values of the KdV  Baker  function  at   branching
points.Note that once again the spectral curve can be  defined  in
terms of Garnier Lax operator
\beq
det(L_{gar}(\lambda)-\rho)=0
\eeq
with
\beq
L_{gar}=x\otimes e_{n+1}-e_{n+1}\otimes  x+{\lambda}^{-1}(x\otimes
x+p\otimes  e_{n+1}+e_{n+1}\otimes  p+x^{2}e_{n+1}\otimes  e_{n+1}
+A) +\lambda B
\eeq
where $e_{i}$ is the standard basis in $R^{n+1}$,
$A=diag(m_{1}^{2},...,m_{n}^{2},0),B=diag(0,..,0,1)$
The nontrivial  tree  amplitudes  at  the  threshold  due  to  the
consideration  above  are  in  one  to  one  correspondence   with
degenerations of the spectral curve.

  It is interesting that the systems  above  allow  quite  different
interpretation in spirit of \cite{gm}as  examples of generic Hitchin
systems on moduli
of  holomorphic  vector  bundles   \cite{hitchin}.The   Lagrangian
providing Hitchin,s dynamics has the following form \cite{gn}
\beq
L=Tr \int_{\Sigma dt}\bar A \dot \phi +{\phi}^{2}+
A_{0}(\bar{\partial}\phi+[\phi,\bar A]+\sum_{i}J_{i}\delta
(z-z_{i}))
\eeq
where $ A,\bar{A}$  are  holomorphic  and  antiholomorphic  G-valued
connections  on  the  surface  $\Sigma$.To  our  purpose   it   is
sufficient to consider  SL(2,C) system  on  the  sphere  which  at  the
quantum       level       generates       Gaudin       Hamiltonian
\cite{nekrasov,rubtsov}.We  are  interested   in   the   classical
dynamics governed by the Hamiltonian $TrL^{2}$ where  L  coincides
with $\phi$ after the resolution  of  the  holomorphic  Gauss  law
constraint,namely for generic case

\begin{displaymath}
L=
\left( \begin{array}{ccc}
a-\sum_{i=1}^{n}\frac{x_{i}y{i}}{\lambda-z_{i}}&b-\sum_{i=1}^{n}
\frac{y_{i}^{2}}{\lambda-z_{i}} \\
c+\sum_{i=1}^{n}\frac{x_{i}^{2}}{\lambda-z_{i}}&-a+\sum_{i=1}^{n}
\frac{x_{i}y_{i}}{\lambda-z_{i}}
\end{array} \right)
\end{displaymath}

According to the  analysis  of
\cite{hah} this Lax operator with  some  additional
constraint  gives   rise to  the
Hamiltonians for Garnier and Neumann systems.For example
the constraints resulting in generic Neumann system looks as
$a=c=0,b=-\frac{1}{2},{\vec x}^{T}\vec x=1,{\vec y}^{T}\vec x=0$
The number  of  marked
points on the sphere corresponds to the number  of  particles  and
positions  of  the  poles  play   the   role   of   the   particle
masses,namely $z_{i}=m_{i}^{2}$.Variables $x_{i},y_{i}$ parametrize
SL(2)  coadjoint  orbit
in the i-th marked point.Finally let us emphasize that the spectral
curve  above  is the cover of the  marked sphere,and wave functions
in the quantum problem
appear to be the four-point conformal blocks at the critical level
in WZW theory.

4.Now we turn to the case when there are two initial particles off
the threshold and the particles on the threshold in the
final state.

Consider first $\phi^{3}$ theory. The equation of motion
on the threshold is
\beq
\label{cubic}
{\partial}_{t}^{2}+m^{2}\phi +{\lambda}{\phi}^{2}=0
\eeq
and        the        asymptotic         condition         becomes
$\phi=a^{*}e^{i\omega t}+O(\lambda)$. The L operator
is the following one
\beq
L= {\partial}_{t}^{2}+m^{2}+2\phi_{cl}
\eeq
Notice that (18) is the (integrated)
first equation in the hierarchy of Novikov's
equations for KDV, it has only one-gap solution. The  L
operator provides three-gap solution in this case
and Baker function necessarily has
three poles on the spectral curve (in the $\omega$-plane)
corresponding to the processes 2 to 1 (forbidden kinematically), 2 to
2, and 2 to 3. Explicitly, the solution of the (\ref{cubic}) with the given
asymptotic condition has the following form
\beq
\phi=\frac{-3}{2\lambda sin^{2}\frac{(t-t_{0})}{2}}
\eeq
where $e^{-it_{0}}=\frac{{\lambda}a^{*}}{6}$
and Baker function

\begin{eqnarray}
\Psi_{k}=e^{-i{\omega}_{k}}(1
-\frac{12}{1-2{\omega}_{k}}\frac{e^{it}}{e^{it}-1}
+\frac{120}{(1-2\omega)(2-2\omega)}(\frac{e^{it}}{e^{it}-1})^{2}\nonumber\\
-\frac{720}{(1-2\omega)(2-2\omega)(3-2\omega)}
(\frac{e^{it}}{e^{it}-1})^{3})
\end{eqnarray}

For the $\phi^{4}$ theory the equation of motion on the threshold
is the first Novikov-type equation for MKDV which again has only
one-gap solution. The potential in the L operator is then the two-gap
one, hence the Baker function has two poles on the spectral curve
corresponding to the processes 2 to 2 and 2 to 4.

For the Sin-Gordon theory the same reasoning gives the Baker function with
only one pole corresponding to 2 to 2 process which is nothing but the
manifestation of the well-known property of factorization in the theory.

Consider now the Standard Model. Reducing onto the threshold we
leave only the Higgs mode $\phi$ thus having the equation
\beq
{\partial_{t}}^{2}{\phi}+\phi+{\lambda}{\phi}({\phi}^{2}-1)=0
\eeq
In the initial state we leave two W-bosons. Thus the field of the
W-boson becomes now the Baker function, the potential in the L
operator being again proportional to the one-gap one. The coefficient
in this case is the ratio of the gauge and the Higgs coupling
constants and when the ratio is equal to N(N+1) the potential is the
N-gap one.
The case with two fermions in the initial state admits a similar
interpretation.

Note also that the problem can be attacked from another point of view.
Indeed,if we assume the mode expansion and take into account only three
modes entering the process then we get (after neglecting the
selfinteraction of the off-threshold modes) an intergrable system of
two degrees of freedom which again is related to the finite-gap solution
of coupled KdV system. But this approach doesn't seem to be the universal one.

5.The picture discussed above admits a generalization to the case
when there are threshold particles both in the initial and final states.
Consider for definiteness the $\phi^{3}$ theory. The appropriate
solution of the equation of motion is now expressed in terms of the Weierstrass
function

\beq
\phi(t)=-\frac{6}{\lambda}(\frac{1}{12}
+\wp(t-t_{0};{\omega},{\omega}^{,}))
\eeq

and the standard parameter $g_{2}=60\acute{\sum} \frac{1}{(2m\omega
 + 2n\acute{\omega})^{4}}$ is equal to $\frac{1}{12}$, while the parameters
$\delta=t_{0}\frac{\pi}{\omega}$ and $q=e^{i\frac{\omega}{\acute{\omega}}}$
should be expressed  in  terms  of  $a$  and $a^{*}$.
It is convenient to use the following representation of the
$\wp$-function
\begin{eqnarray}
\wp(u)=-\frac{1}{12}(\frac{\pi}{\omega})^{2}+(\frac{\pi}{\omega})^{2}
\sum_{m=1}\frac{2mq^{2m}}{1-q^{2m}}-\nonumber\\
(\frac{\pi}{\omega})^{2}\sum_{m=1}
me^{-im\frac{{\pi}u}{\omega}}\frac{q^{2m}}{1-q^{2m}}-
(\frac{\pi}{\omega})^{2}\sum_{m=1}
me^{+im\frac{{\pi}u}{\omega}}\frac{1}{1-q^{2m}}
\end{eqnarray}
Comparing it with the asymptotic condition one sees that
unlike the case with the threshold particles in the final state only
the parameters of the classical solution are fixed only in the leading in
$\lambda$ approximation:
\begin{eqnarray}
q^{2}=(\frac{\lambda}{6})^{2}aa^{*}(1+o(\lambda)) \\
\omega=\pi+o(\lambda)                                \\
e^{-2i\delta}={\frac{a}{a^{*}}}q^{2}
\end{eqnarray}
Another difficulty seems to be related to the fact that the higher order terms
in $\omega$ introduce polynomial dependence on time after expansion of
the classical solution in $a$ and $a^{*}$. Both these problems are
related to the fact that the  kinematics considered admits resonant
amplitudes and the naive perturbation expansion of the  is not
correct in this case. We consider these problems and describe the
corresponding
nullification phenomena elsewhere \cite{gs}.

Note that  some important questions remain beyond the scope of the paper.
At first one can wonder if higher times can be included into the game.The
important step in this direction was made in \cite{Makeenko} where it was
shown that in the large N limit of O(N) model Shwinger-Dyson equation
can be solved exactly and the solution implies that the role of the third
KdV time is played by the coupling constant.Another line of reasoning
suggests the nontrivial space dependence of the singularity surface which
results in some quasiclassical configuration responsible for the threshold
amplitudes (see reviews \cite{rubakov2} and references therein).The meaning
of these quasiclassical configurations (for instance quantized bubbles from
\cite{gv}) in the integrability framework is currently unclear. Mention
also a more formal problem of generalization of the Yang-Baxter equation
to the case when there is a finite number of nondiagonal matrix elements in
the S-matrix.

The work was partially supported by ISF grant MET300, by INTAS
grant 1010-CT-93-0023 and by RFFI grant 95-01-01101. K.S. was also partially
supported by INTAS grant 93-633. One of us (K.S.)
thanks A.Rosly for the useful discussions.

\end{document}